# New Survey Questions and Estimators for Network Clustering with Respondent-Driven Sampling Data

*Ashton M. Verdery[†1], Jacob C. Fisher[2], Nalyn Siripong[3], Kahina Abdesselam[4], Shawn Bauldry[5]*

[†] Corresponding author;

(1) The Pennsylvania State University; (2) Duke University; (3) University of North Carolina – Chapel Hill; (4) University of Ottawa; (5) Purdue University

**Main text word count: 7873**

**Abstract**

Respondent-driven sampling (RDS) is a popular method for sampling hard-to-survey populations that leverages social network connections through peer recruitment. While RDS is most frequently applied to estimate the prevalence of infections and risk behaviors of interest to public health, like HIV/AIDS or condom use, it is rarely used to draw inferences about the structural properties of social networks among such populations because it does not typically collect the necessary data. Drawing on recent advances in computer science, we introduce a set of data collection instruments and RDS estimators for network clustering, an important topological property that has been linked to a network's potential for diffusion of information, disease, and health behaviors. We use simulations to explore how these estimators, originally developed for random walk samples of computer networks, perform when applied to RDS samples with characteristics encountered in realistic field settings that depart from random walks. In particular, we explore the effects of multiple seeds, without vs. with replacement, branching chains, imperfect response rates, preferential recruitment, and misreporting of ties. We find that clustering coefficient estimators retain desirable properties in RDS samples. This paper takes an important step towards calculating network characteristics using non-traditional sampling methods, and it expands RDS's potential to tell researchers more about hidden populations and the social factors driving disease prevalence.

**Acknowledgements**: We thank M. Giovanna Merli, Ann Jolly, and Anne DeLessio-Parson for providing information about aspects of the empirical cases we examine. We acknowledge assistance provided by the Population Research Institute, which is supported by an infrastructure grant by the Eunice Kennedy Shriver National Institute of Child Health and Human Development (R24-HD041025), and from a seed grant provided by the Institute for CyberScience at Penn State University. Portions of this research were funded by NCHS grant 1R03SH000056-01 (Verdery PI).

**Introduction**

Researchers in many fields are interested in populations that cannot be sampled by conventional methods because they are rare, lack a sampling frame, or unwilling to participate in traditional survey protocols. Such groups, known as hidden populations (Heckathorn 1997), are often marginalized and at high risk of infections like HIV/AIDS. Respondent-Driven Sampling (RDS) is a set of methods for sampling and making inferences about hidden populations that has proliferated throughout the social sciences and public health (Malekinejad et al. 2008; White et al. 2012). RDS uses a without-replacement "link-tracing" approach, similar to snowball sampling, where each respondent attempts to recruit a limited number of her personal network contacts in the target population until the desired sample size is attained. RDS offers a popular, quick, cost-effective, and anonymous approach for sampling understudied groups like the homeless, drug users, or commercial sex workers that claims to provide asymptotically unbiased estimates of the population mean under limited conditions (Volz and Heckathorn 2008; Salganik and Heckathorn 2004). Though concerns exist about RDS's validity (Gile and Handcock 2010; Verdery et al. 2016; Merli, Moody, Smith, et al. 2015; Lu et al. 2013; Lu et al. 2012; Goel and Salganik 2010; Tomas and Gile 2011; McCreesh et al. 2012; Fisher and Merli 2014; Crawford et al. 2015), continued development of estimators, diagnostics, and reporting protocols are increasing its legitimacy (Lu 2013; Verdery et al. 2015; Gile 2011; Gile and Handcock 2011; Gile, Johnston, and Salganik 2015; White et al. 2015; Nesterko and Blitzstein 2015; Yamanis et al. 2013; McCreesh et al. 2013; Crawford 2016).

Most RDS studies focus on prevalence estimation – that is, estimation of the population mean or proportion of a focal attribute like condom use – and avoid making inferences about other relevant estimands. We focus on network structure and, in particular, clustering. The structure of both social and contact networks is a key component of the risk environment for members of hidden populations (Rhodes and Simic 2005) with important implications for disease transmission (Schneider et al. 2012; Morris et al. 2009) and health behaviors (Centola and Macy 2007). Highly clustered risk networks, like sexual contact networks or shared needle networks, can lead to more redundant paths, making disease transmission more likely (Moody 2002) and altering the relationship between concurrency and epidemic potential (Moody and Benton 2016). Clustering can also have benefits. Highly clustered friendship networks lead to normative reinforcement, and can increase individual likelihoods of engaging in and spreading health-promoting behaviors like joining an internet-based health forum (Centola 2010), adopting modern contraceptives (Kohler, Behrman, and Watkins 2001), abstaining from illicit drugs (Silverman et al. 2007), getting tested for HIV (Karim et al. 2008), or avoiding unprotected sex (Lippman et al. 2010). Normative reinforcement through clustering can also drive unhealthy behaviors, such as sexual concurrency (Yamanis et al. 2015).

Despite its sociological and epidemiological importance, few studies of hidden populations using RDS have directly examined network structure. This is by design: because field implementations of RDS require that samples be conducted *without* replacement and with maximal anonymity, typical RDS samples have limited opportunity to measure network structure beyond recruiter-recruit relationships. Some have proposed using RDS to measure homophily (Wejnert 2010), or the tendency for people with similar attributes to be tied (McPherson, Smith-Lovin, and Cook 2001), but these approaches are flawed (Crawford et al. 2015) and there have been few developments since. Others have fit exponential random graph models to RDS data (Merli, Moody, Smith, et al. 2015; Gile and Handcock 2011), but learning about networks themselves was not the primary purpose of these studies. The ability of RDS studies to estimate network structure is important, however, because without closer attention to network characteristics that influence risk behaviors and sexually transmitted infections, RDS studies will be unable to offer a comprehensive picture of the dynamics driving epidemic transmission or other network diffusion processes.

This paper focuses on the performance of recently developed estimators of network clustering that can be applied to RDS data. Work in computer science has proposed clustering estimators for data obtained via random walk sampling (RWS) (Hardiman and Katzir 2013), which is an alternative link-tracing sampling design more appropriate for computer networks than human populations. RDS procedures depart from RWS in several important ways that call into question whether such estimators can apply to RDS. We interrogate this question throughout the paper. First we discuss measures of network clustering, then we introduce their estimation in network censuses vs. samples and review how RDS differs from RWS. Throughout, we focus on RDS data collection strategies that could inform clustering estimators, which leads us to introduce two alternative survey question approaches for RDS. We next use simulation procedures to evaluate whether our proposed survey questions and estimators of network clustering are appropriate for RDS data, focusing on bias, sampling variance, and total error. We then discuss how our proposed survey questions perform in 6 empirical RDS surveys. Our results indicate that the estimators maintain reasonable properties with RDS data and that the questions have good empirical properties. These findings lead us to suggest that researchers add clustering questions and estimators to RDS protocols to further explore network structure. We conclude by focusing on the potential benefits of clustering estimation with RDS data.

**Background**

*Initial Notation*

The following notation guides our discussion throughout the paper. For illustrative purposes, we rely on Figure 1, which shows a) a hypothetical population (i.e., nodes A through I); b) the social network linking its members (solid lines connecting nodes); c) a hypothetical time-ordered RWS link-tracing sample starting from node A (dashed, directed, and numbered lines); and d) a table counting relevant nodal statistics shown (on the right). Note that item (c) refers to a random walk sample (RWS) rather than a respondent-driven sample (RDS); in an RDS sample, node E would be ineligible to be sampled a second time because RDS is conducted *without* replacement. Below, we review this and other differences between RWS and RDS that together call into question whether clustering estimators designed for RWS can be applied to RDS samples.

**Figure 1. Example network with a hypothetical random walk sampling (RWS) sample and components needed to calculate local and global clustering coefficients for the whole network.**

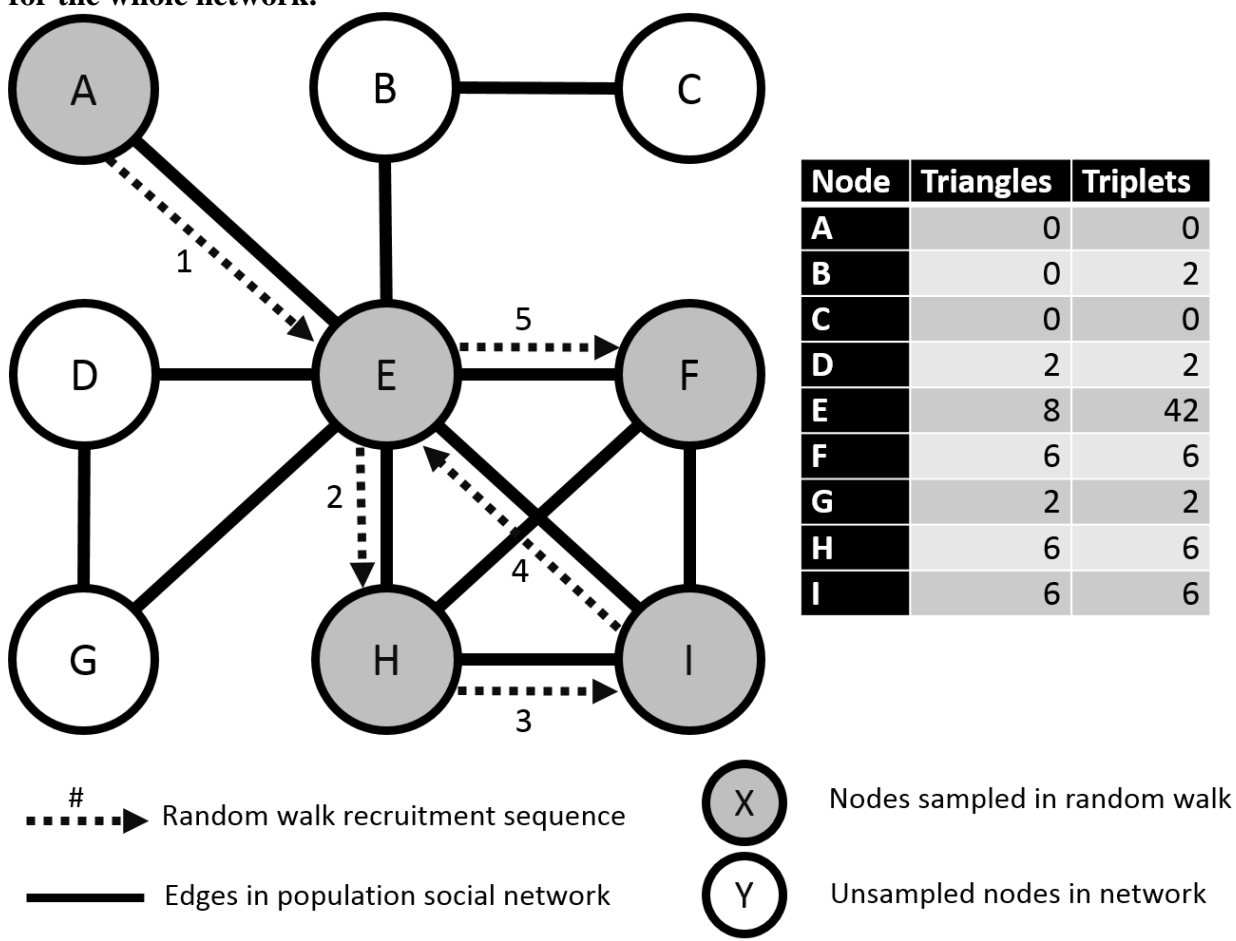

We characterize a social network of $n$ people as a graph $G$ with nodes $V$ representing people and undirected edges $E$ representing social ties. In Figure 1, we label nodes A through I and represent edges as undirected solid lines (we discuss the time-ordered, directed random walk steps shown with dashed and numbered lines below). We represent the graph as an $n \times n$ adjacency matrix, $A$, whose elements, $a_{ij}$ are 1 if there is a tie (edge) from person $i$ to person $j$ (i.e., when $i \leftrightarrow j$) and are 0 otherwise. For instance, there is an edge between nodes B and C in Figure 1 (but not between nodes A and B). We follow standard practices in the RWS and RDS literatures (Lovász 1993; Hardiman and Katzir 2013; Volz and Heckathorn 2008) and consider an undirected graph with one component (see Lu et al. 2013 for the performance of RDS in directed networks). Since the network is undirected, the adjacency matrix $A$ is symmetric and $a_{ij} = a_{ji}$ for all $i = 1, \dots, n$ and $j = 1, \dots, n$. We set the diagonal of $A$ to 0 (i.e., $a_{ii} = 0$ for all $i = 1, \dots, n$).

For convenience, we define $d_i = \sum_{i=1}^{n} a_{ij} = \sum_{j=1}^{n} a_{ij}$ as the *degree* of person $i$, meaning how many ties $i$ has in the network. In figure 1, node A's degree is 1 because he or she is only linked to one other node (E), while node B's degree is 2 because he or she is linked to both E and C. In empirical RDS studies, researchers typically estimate degree by asking respondents questions like "how many people do you know (you know their name and they know yours) who have exchanged sex for money in the past six months?" (WHO 2013, 147). Some have studied the effect of inaccurate degree reporting on RDS estimates (Neely 2009; Lu 2013; Lu et al. 2012), but we assume accurate degree reporting.

*Clustering coefficients*

Watts and Strogatz (1998) introduced the clustering coefficient to characterize small world networks (Milgram 1967). Small world networks are a) highly clustered, meaning most ties between people appear in pockets of interconnection (see below), and b) have short average path lengths, meaning that the minimum number of steps between network members is, on average, low (e.g., as embodied in the famous phrase "six-degrees of separation"). Clustering coefficients measure the first criterion.

Watts and Strogatz originally proposed a global measure of the clustering coefficient, defined as

$$GCC = \frac{2\sum_{i=1}^{n}\sum_{j=1}^{n}\sum_{k=1}^{j-1} a_{ij}a_{ik}a_{jk}}{\sum_{i=1}^{n} d_i(d_i-1)} \quad (1)$$

where i, j, and k index unique respondents (Hardiman and Katzir 2013; Newman, Strogatz, and Watts 2001; Watts and Strogatz 1998). The global clustering coefficient (GCC) summarizes the overall network clustering by dividing the count of triangles by the count of connected triplets, where triangles are defined as sets of three individuals ($i$, $j$, and $k$) for whom cells $a_{ij}$, $a_{ik}$, and $a_{jk}$ in the adjacency matrix $A$ are all equal to 1 and connected triplets are defined as sets of three individuals ($i$, $j$, and $k$) where cells $a_{ij}$ and $a_{ik}$ are equal to 1. As such, triangles are a subset of connected triplets, ones with a connection in cell $a_{jk}$. A node's number of connected triplets is a function of his or her degree, i.e., node $i$'s number of connected triplets is $d_i(d_i - 1)$. Figure 1's embedded table holds triangle and connected triplet counts for each node. The GCC of this graph is $30/66 = 0.4545$. It is important to note that equation (1) cannot be evaluated for most RDS studies without information on connections between unsampled peers. We introduce simple questions for RDS surveys that address this issue below.

Extensions to the clustering coefficient concept consider the average amount of clustering among each individual's affiliates in the network. This second measure, the local clustering coefficient (LCC), is defined as

$$C_{LCC} = n^{-1} \sum_{i=1}^{n} \frac{2\sum_{j=1}^{n}\sum_{k=1}^{j-1} a_{ij}a_{ik}a_{jk}}{d_i(d_i-1)}. \quad (2)$$

The LCC measures the average of each individual's number of triangles divided by his or her connected triplets. In Figure 1, the LCC is obtained by first dividing triangles by connected triplets, then taking the average (when $d_i = 1$, the value is set to 0). Thus, nodes A-C each contribute values of 0 to the LCC, while node D contributes a value of $0.111 = 2/2 * 1/9$ and node E contributes a value of $0.278 = 8/32 * 1/9$, and so on. This graph's LCC is 0.5767. As with the GCC, the LCC cannot readily be evaluated for many RDS samples. The key difference between the clustering coefficient measures is that the GCC captures the totality of network members' experience, which may be dominated by low clustering among high degree nodes, for instance, while the LCC captures the average experience of network members, where each person in the network is weighted equally.

Although clustering coefficients are recent additions to the social networks literature, they resemble other important network characteristics, in particular, transitivity, ego-network density, and measures of clustering from the exponential random graph modeling framework. We omit detailed discussion of these alternate measures for the sake of brevity.

*Measuring Clustering in Network Censuses and Samples*

The calculation of many network-level statistics, including the clustering coefficient, assumes that researchers measure the entire adjacency matrix, $A$, in terms of cells (edges) and rows/columns (nodes). In Figure 1, it would be assumed that the researcher measured all ties (solid, undirected lines) and nodes (labeled A-I). Collecting such saturated network data is challenging (Smith 2012), however, and often impossible for populations without clearly defined institutional boundaries (e.g., schools). In other settings, either intentionally or not, researchers do not collect data on all network members (node missingness), do not measure all relevant ties linking network members (edge missingness), or both.

When researchers cannot conduct a census of the network, they often turn to samples. There are many approaches to collecting sampled network data, including randomly drawn samples (Marsden 1987; Krivitsky, Handcock, and Morris 2011; Smith 2012; McPherson, Smith-Lovin, and Brashears 2006) and numerous link-tracing approaches (Goodman 1961; Heckathorn 1997; Volz and Heckathorn 2008; Mouw and Verdery 2012). We focus on the latter.

*Hardiman and Katzir Estimators*

Hardiman and Katzir (2013) introduce estimators for the LCC and GCC that use data gathered in an RWS sample, like that shown in Figure 1. Intuitively, for vertices $x_1, x_2, \ldots, x_r$ sampled via RWS, they estimate clustering with the presence of a tie between the vertices before and after the focal vertex. Typical RDS studies do not ask about the existence of this tie, though some have (see application section below and Appendix B), and in the next section we propose two question formats for RDS studies to assess its existence. More formally, for a step $k$ in a random walk, $X$, let $\phi_k$ represent whether a tie is present between the vertex before $x_k$, i.e., $x_{k-1}$, and the vertex after $x_k$, i.e., $x_{k+1}$. In the random walk depicted in Figure 1, for instance, $\phi_k$ would be 0 the first time node E is sampled because nodes A and H are unconnected, but it would be 1 the second time node E is sampled because nodes F and I are connected. That is, $\phi_k = a_{(x_{k-1}, x_{k+1})}$ for each $2 \leq k \leq r-1$, where $a_{ij}$ is the cell in the $i^{\text{th}}$ row and the $j^{\text{th}}$ column of the adjacency matrix, as before. Importantly, $\phi_k$ is not calculated for the first and last nodes of the walk, because the former has no recruiter and the latter no recruitee.

Next for the LCC, define a weighted sum of the $\phi$ value as $\mathbf{\Phi}_l = \left(\frac{1}{r-2}\right) \sum_{k=2}^{r-1} \phi_k \left(\frac{1}{d_{x_k}-1}\right)$. In this case, $d_{x_k}$ represents the degree of the vertex $x_k$ in the random walk and $r$ is the length of the random walk. Thus, $\mathbf{\Phi}_l$ is the average of whether the previous vertex in the random walk ($x_{k-1}$) and the following vertex in the random walk ($x_{k+1}$) were tied, weighted by the probability of observing the current vertex. In RWS on an undirected, unweighted graph, the probability of observing a given vertex is the inverse of that vertex's degree if the random walk is in the steady state, which is typically achieved if the walk is sufficiently long or started with steady state probabilities (reviewed in greater depth in Verdery et al. 2016; Lovász 1993). We note that this finding cannot be assumed to hold for the finite, branching, without replacement samples conducted in RDS and future research may investigate alternate weighting schemes.

Finally, let $\mathbf{\Psi}_l = (1/r) \sum_{k=1}^{r} (1/d_{x_k})$, representing the sum of sampled vertices' reciprocal degrees. Hardiman and Katzir define an estimator of the LCC as:

$$\hat{c}_{LCC} = \frac{\mathbf{\Phi}_l}{\mathbf{\Psi}_l} = \frac{\left(\frac{1}{r-2}\right) \sum_{k=2}^{r-1} \phi_k \left(\frac{1}{d_{x_k}-1}\right)}{\left(\frac{1}{r}\right) \sum_{k=1}^{r} \left(\frac{1}{d_{x_k}}\right)} \qquad (3)$$

Hardiman and Katzir also develop an estimator of the GCC. Letting $\mathbf{\Phi}_g = (1/(r-2)) \sum_{k=2}^{r-1} \phi_k d_k$ and $\mathbf{\Psi}_g = (1/r) \sum_{k=1}^{r} d_{x_k} - 1$, they suggest the following measure for the global clustering coefficient:

$$\hat{c}_{GCC} = \frac{\mathbf{\Phi}_g}{\mathbf{\Psi}_g} = \frac{\left(\frac{1}{r-2}\right)\sum_{k=2}^{r-1} \phi_k d_{x_k}}{\left(\frac{1}{r}\right)\sum_{k=1}^{r} d_{x_k} - 1} \qquad (4)$$

Hardiman and Katzir use both analytic proofs and simulation to show that their proposed estimators are asymptotically unbiased with minimal variance for large RWS samples and that they produce more consistent results at any given sample size than other approaches that query each sampled node's full ego network (counting ego network reports in the sample size). Although RDS does not rely on simple random walks, researchers may wish to apply these estimators to RDS samples. The following section discusses RDS departures from RWS with special attention to the empirical contexts in which RDS studies are conducted. Within it, we propose new survey questions that researchers could employ to estimate clustering via the Hardiman and Katzir estimators. We examine how these questions perform in six empirical surveys in the discussion section.

*RDS Departures from RWS*

The Hardiman and Katzir estimators cannot immediately be applied to RDS studies in the field because they were developed for RWS, which differs considerably in core assumptions. Deviations of RDS from RWS have been shown in prior work to bias other estimators, like that of the population mean (Gile 2011; Merli, Moody, Smith, et al. 2015; Tomas and Gile 2011) and sampling variance (Verdery et al. 2016), so we should not expect that a naïve application of Hardiman and Katzir's clustering coefficient estimators will yield viable estimates from empirical RDS samples.

**Table 1. Comparison of features of RWS and RDS.**

|  | RWS | RDS |
|---|---|---|
| 1) Number of seeds | One | Multiple |
| 2) Seed selection | Proportional to steady state | Convenience |
| 3) Branching | No | Yes |
| 4) Replacement | Yes | No |
| 5) Link tracing efficacy | 100% | Less than 100% |
| 6) Preferential recruitment | No, researcher controls | Yes, respondent controls |
| 7) Sample size | Large (more than 10,000) | Small (less than 1,000) |
| 8) Measurement of $\phi_k$ | Can be queried | Asked of respondent |

Table 1 summarizes 8 RDS departures from RWS that may affect clustering estimation. A RWS sample of a network begins with selecting a single "seed" node, typically with probability proportionate to the steady state probability, $\pi_i = d_i/2m$, where $d_i$ is the degree of node $i$ in the population and $m = 1/2 \sum_i d_i$ is the number of edges in the population (Lovász 1993). By contrast, most RDS protocols recommend initiating the sample by identifying, often by convenience, eight to ten members of the hidden population who are willing to participate, have large personal networks with other members of the target population, and are diverse with respect to relevant focal attributes, such as years injecting drugs (WHO 2013, 71–82). A first consequence of this distinction is that RWS samples lead to a single chain in a network (as in the hypothetical chain depicted in Figure 1), whereas RDS samples start from multiple points and yield multiple chains. A second consequence is that RDS samples often exhibit seed dependence, whereas RWS samples do not (Gile and Handcock 2010).

RWS and RDS also differ in their approach to tracing links. RWS samples proceed without branching (i.e., one coupon), while RDS almost always allow branching in practice through the distribution of two or three recruiting coupons to each respondent (Goel and Salganik 2009). RWS samples are conducted with replacement while RDS is conducted without replacement, which means that recruitment becomes competitive (Heckathorn 1997; Barash et al. 2016; Gile and Handcock 2010; Gile 2011; Crawford 2016). Other differences arise because RWS is researcher-driven (or algorithm-driven), while RDS is respondent-driven. In RDS, respondents must identify, approach, and successfully recruit peers, which can yield less than perfect link tracing efficacy and introduce preferential recruitment (Merli, Moody, Smith, et al. 2015; Verdery et al. 2015).

Sample size is another distinction because RWS samples are used in computer science or fields where costs of sampling additional individuals is low compared to RDS in human populations (Mouw and Verdery 2012) . For instance, Hardiman and Katzir examine their estimators performance in four large networks with 1% samples of sizes $n = 9,780$, $n = 21,734$, $n = 30,724$, and $n = 48,440$. By contrast, Malekinejad et al. (2008) report attained sample sizes for 63 RDS studies, ranging from $n = 99$ to $n = 548$, with a median $n = 152$. A first consequence of smaller samples is that RDS samples are more likely to contain finite sampling bias even when assumptions are met because the samples are too small for asymptotically unbiased RDS estimators to minimize bias. A second consequence of small RDS samples is that they are likely to violate the RDS assumption that the sample is "in equilibrium", a fact exacerbated by convenience sampling seeds (Gile and Handcock 2010; Wejnert 2009).

The final departure of RDS from RWS is anonymity, which pertains to the measurement of $\phi_k$, whether person $x$'s recruiter knows person $x$'s recruitee. First, unlike in computer or online networks where it is comparatively easy to determine for each node $x_k$ in the random walk, whether the prior node, $x_{k-1}$, is tied to the subsequent node, $x_{k+1}$, this task is more challenging in an RDS sample of a human population. One cannot seek $x_{k-1}$ in a stored contact list of node $x_{k+1}$ or otherwise backtrack the sample for direct measurement; rather, the existence of this tie must be elicited from respondents themselves during a period when the respondent is answering the survey, which can introduce measurement error and other challenges. The timing of recruitments and preservation of anonymity in RDS mean that a) researchers cannot ask about recruitments that have not yet occurred (e.g., cannot ask A whether he or she is tied to H in the RWS in Figure 1), and b) researchers cannot divulge who recruited whom to respondents (e.g., cannot tell H that A recruited E). The middle recruit is the only feasible person to ask about this tie's existence in an RDS sample (E in this example), although this requires E to report on a tie that exists between two of his alters and thus may introduce reporting error (a topic we examine below).

In many RDS surveys, a majority of respondents participate twice, once when they are recruited themselves (primary interview) and a second time when they return to the research site to collect additional incentives for successfully recruiting peers (secondary interview). Acknowledging this interview timing, we propose two questions that researchers can ask RDS respondents to feasibly elicit information about potential ties between $x_{k-1}$ and $x_{k+1}$:

A) **[In the secondary interview]**. "Does the person who gave you the coupon know the person who you gave the coupon to or vice versa." (We refer to this from here on as the *binary question* format).

B) **[In the primary or secondary interview]**. "What percent of people who you know in the population does the person who gave you the coupon know." (We refer to this from here on as the *percentage question* format).[1]

The binary question format garners the exact information required by the Hardiman and Katzir estimators, but it relies on the accuracy of respondent reports about recruiter-recruitee relationships. It also can only be estimated on a subset of sampled cases, as it cannot be asked until the secondary interview (after recruitment). The percentage question format differs from Hardiman and Katzir's suggested approach, but it can be asked during either the main survey (of all respondents) or the follow-up interview (of the subset of respondents who recruit). If asked in both, researchers can check test-retest validity and potentially diagnose respondent comprehension problems. Of course, there are other possible ways to ask such questions in RDS surveys, but our proposed approaches are flexible in terms of implementation and preserve the desirable confidentiality of standard RDS studies.

---

[1] Note, many studies do not ask respondents directly for the percentage. Rather, they ask them to report personal network size (e.g., "A1. How many adult sex workers do you know who live in this city?"), then to report the number known by the recruiter (e.g., "A2. Of the number in A1, how many are known by the person who gave you the coupon?"). Percentages can be calculated directly from this pair of questions. We review six surveys that asked variants of the questions needed to calculate the clustering coefficient estimators in Appendix B.

**Data and Methods**

*Approach*

We first evaluate the performance of Hardiman and Katzir's estimators applied to RDS through simulation methods. We aim to understand the effects of increasingly large departures from RWS, toward more realistic situations encountered within RDS data collection. To do this, we simulate data collection from underlying population social networks. It is notoriously difficult to obtain analytical results for RDS estimators, which is why many prior developments have tested proposed estimators through simulation. We test scenarios driven by data collection parameters to match how RDS departs from RWS, drawing 1000 samples in each scenario. It is important to draw multiple samples per scenario to determine the estimators' distributional properties (bias, sampling variance, and total error). For each simulated sample, we calculate the Hardiman and Katzir LCC and GCC estimators implemented with both question formats we proposed. We compare these sample estimates to the parameters in the population social network (or as would be calculated in a census). After examining how Hardiman and Katzir's estimators perform in simulations, we evaluate their feasibility in six empirical RDS samples.

*Data*

**Figure 2. Largest Weakly Connected Component of Project 90 Data Set, Nodes Colored by Race (Grey = White; Black = Non-White) and Sized by Degree. The network is displayed using the ForceAtlas2 algorithm, with no node overlap, in Gephi 0.9.**

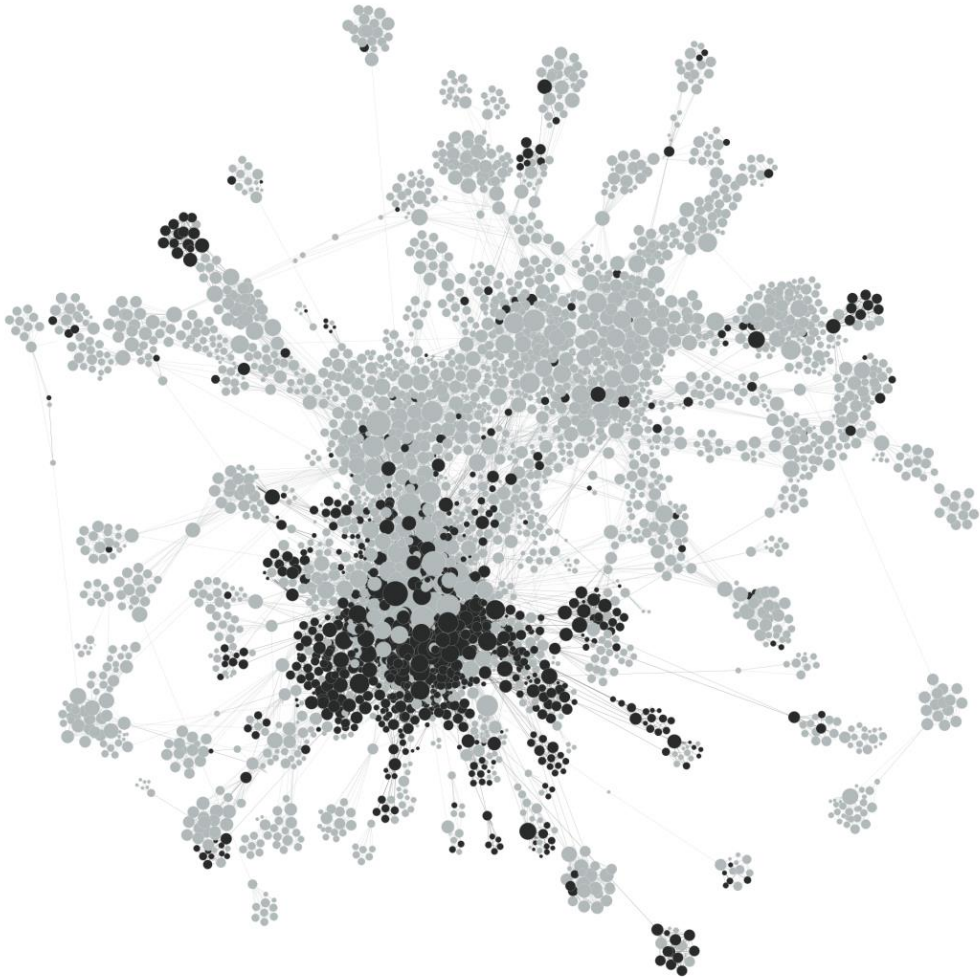

We first simulate link-tracing samples from a hidden population social network of heterosexuals, sex workers, and injecting drug users at elevated risk of HIV/AIDS collected beginning in 1987 as part of the Project 90 study in Colorado Springs, CO (Potterat et al. 2004; Woodhouse et al. 1994; Rothenberg et al. 1995; Klovdahl et al.

1994). The project aimed to assess how network structure affected disease transmission, and, as such, the researchers sought to obtain a census of the hidden population and their links to one another. These data have previously been used in prior RDS assessments (Goel and Salganik 2010) and are made available to researchers by the Office of Population Research at Princeton University ("Office of Population Research, Princeton University" 2015). We focus on 4,111 individuals linked by 17,164 ties that remain in the network's largest weakly connected component after dropping cases lacking valid attribute codes. Figure 2 shows the network linking members of this population, with nodes colored by a key structuring variable (white/non-white). Whites make up 74.7% of network members, while 20.6% of ties cross race categories. Nodes of different races group together in different parts of the figure, but there are many cross group links.

**Table 2. Summary network statistics for data sets analyzed in this paper.**

| Network | Nodes | Edges | Density | GCC | LCC | Cross group ties** |
|---|---|---|---|---|---|---|
| Project 90 | 4,111 | 34,328 | 0.002 | 0.657 | 0.348 | 0.206 |
| Facebook Nets* | | | | | | |
| Minimum | 4,985 | 212,114 | 0.004 | 0.200 | 0.135 | 0.015 |
| 25th Percentile | 5,930 | 367,486 | 0.008 | 0.216 | 0.152 | 0.032 |
| Median | 6,877 | 503,939 | 0.013 | 0.231 | 0.167 | 0.038 |
| 75th Percentile | 7,840 | 705,501 | 0.014 | 0.241 | 0.179 | 0.054 |
| Maximum | 9,693 | 905,428 | 0.017 | 0.276 | 0.199 | 0.163 |

**Notes: *Statistics presented for the Facebook networks are computed separately, the largest network does not necessarily have the largest proportion of cross group ties, for instance; **Cross group ties refer to ties that cross white/non-white categories in Project 90 and ties that cross freshmen/non-freshmen categories in the Facebook networks.**

To understand how the Hardiman and Katzir estimators perform across a range of networks, we also examine additional networks from a data set of 100 Facebook networks collected in 2005, which have also been subject to intensive examinations in prior simulation evaluations of RDS (Mouw and Verdery 2012; Verdery et al. 2016). Importantly, because they were collected when Facebook was new and membership restricted to those with college email addresses, researchers have argued that these networks represent realistic, offline social and interaction networks (Traud, Mucha, and Porter 2012; Traud et al. 2011; Clouston et al. 2009). We restrict analysis to 29 university networks where the largest connected component of users with valid attribute codes contained between 5,000 and 10,000 nodes, size restrictions we put in place to avoid without replacement sampling effects (Barash et al. 2016) and to maintain computational tractability. The Project 90 network is smaller, less dense, more clustered, and less homophilous than the Facebook networks.

*Scenarios*

We provide a replication file for researchers interested in replicating and expanding our scenarios for the Project 90 network, which is publicly available data. In both data sets, we focus on five scenarios designed to test the bias, sampling variance, and error of Hardiman and Katzir's estimators when used with standard RDS protocols as opposed to simple RWS. Table 3 shows what key features we manipulate in each scenario. We first simulate collecting simple random walks ("RWS baseline"). These scenarios begin from a single seed selected with steady state probabilities, are conducted *with* replacement, do not branch, experience 100% link-tracing efficacy without preferential recruitment, and do not contain any measurement error for $\phi_k$.

We then selectively relax parameters until the samples resembles the standard RDS protocol. We start with a scenario designed to mimic an ideal case of RDS constrained by the method's actual implementation in the field ("RDS baseline"). This scenario's samples begin from 10 seeds selected via convenience sampling (implemented as uniform random seed selection in the main text; in appendix A we consider four other seed selection scenarios and find that they did not alter our results), are conducted *without* replacement (recruitment is competitive between respondents), and may branch up to three ways from each respondent (i.e., each respondent is simulated as having 3 coupons), respondents always approach and succeed in recruiting peers who have not already been sampled (i.e., 100% recruitment efficacy), selecting them at random among the set of their friends who have not participated (no preferences), and respondents accurately report the items used to measure $\phi_k$ (either the presence or absence of a tie between their recruiter and their recruitee for the binary question format, or the percentage of their potential recruitees known by their recruiter). This RDS baseline scenario subsumes the first four ways that RDS departs from RWS listed in Table 1.

We next examine the fifth through seventh ways that RDS departs from RWS. We look at how less than perfect recruitment efficacy affects estimates by considering a scenario where only 80% of offered coupons are accepted by the targeted peer ("+ less than 100% efficacy"). We then test effects of preferential recruitment ("+ preferential recruitment"), modeling it as a case where all respondents are half as likely to offer to certain types of peers (to white peers in the Project 90 network and freshmen in the Facebook networks). Finally, we examine what happens when respondents misreport recruiter-recruitee ties ("+$\phi_k$ measurement error"). For the binary question format where respondents report on the presence or absence of a tie between their recruiter and recruitee, we subject each report to a 10% random chance of being misattributed (ties reported as non-ties or non-ties reported as ties). For the percent question format where respondents report on the percent of their network alters known by their recruiter, we randomly shift this number by up to ±10% from its true value (capping responses at 0 or 1).

**Table 3. Parameters used in each simulation scenario.**

| Scenario | Seeds | Selection | Replace | Branches | Efficacy | Preferential | Error |
|---|---|---|---|---|---|---|---|
| RWS baseline | 1 | Steady state | Yes | 1 | 100% | No | 0% |
| RDS baseline | 10 | Convenience | No | 3 | 100% | No | 0% |
| +imperfect (80% efficacy) | 10 | Convenience | No | 3 | 80% | No | 0% |
| +preferences (targeted recruitment) | 10 | Convenience | No | 3 | 80% | Yes | 0% |
| +misreporting ($\phi_k$ mismeasurement) | 10 | Convenience | No | 3 | 80% | Yes | 10% |

In all simulated samples we assume respondents accurately report degree. Although sample size marks a key way in which RDS departs from RWS, we hold target sample sizes constant at 400, which is a small fraction of the population sizes we examine. We found that target sample sizes were attained in all scenarios, which reviews of RDS indicate happens frequently (Malekinejad et al. 2008).

*Measures*

We measure the performance of Hardiman and Katzir's clustering coefficient estimators with three indicators. For each of the question formats (binary or percentage) of each of the estimators (GCC or LCC) in each scenario, we calculate a) their bias, defined as $bias = a^{-1} \sum_{i=1}^{i=a}(\hat{c} - C)$ where $a$ is the number of simulated samples; b) their sampling variance (SV), defined as $SV = a^{-1} \sum_{i=1}^{i=a}(\hat{c}_i - a^{-1} \sum_{j=1}^{j=a} \hat{c}_j)^2$; and c) their root mean square error (RMSE), defined as $RMSE = \sqrt{(bias^2 + SV)}$.

**Simulation Results**

**Figure 3. Performance of Hardiman Katzir estimators by estimator and question format in RWS on the Project 90 data set.**

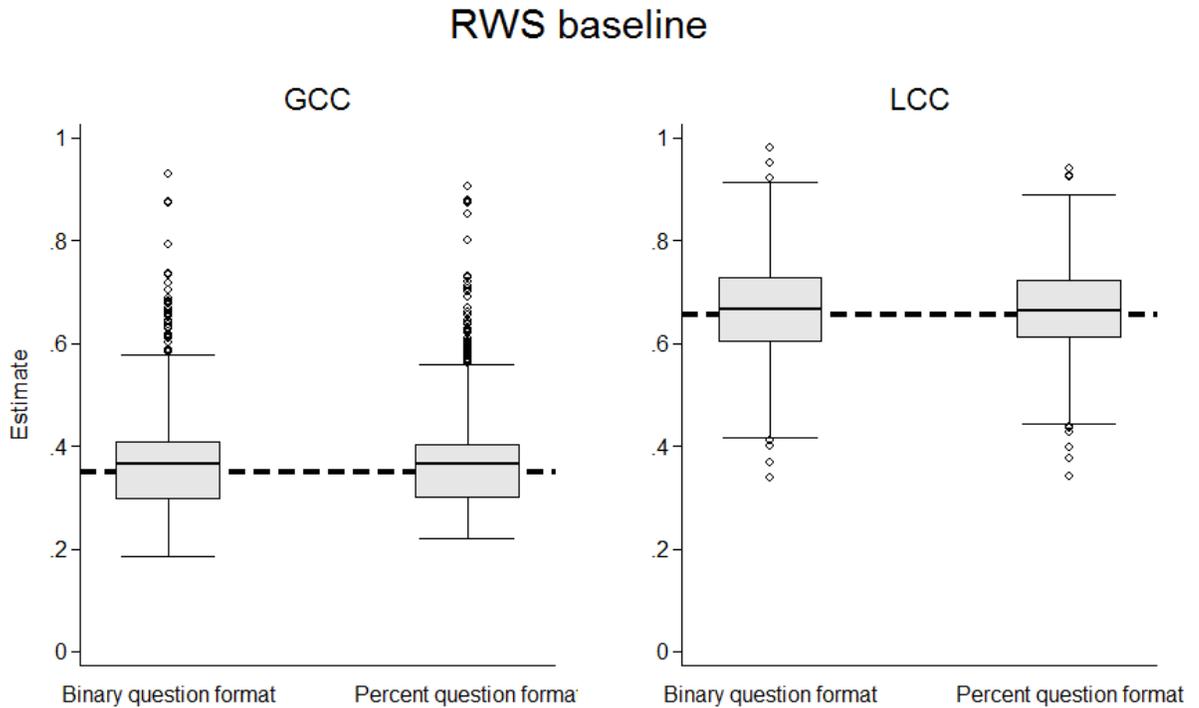

Note: these are non-standard box plots that show the mean rather than the median as the central line; thick dashed line indicates the population parameter

We first consider the distribution of estimates for both the GCC and LCC calculated via the binary and percent question formats in the baseline RWS scenario on the Project 90 network. Figure 3 shows that both estimators, using either question format, exhibit minimal bias that arises because of finite sample sizes. The LCC estimator is less biased than the GCC estimator ($GCC\ binary\ bias = 0.017; LCC\ binary\ bias = 0.009; GCC\ percent\ bias = 0.017; LCC\ percent\ bias = 0.008$). Sampling variance is approximately equivalent across estimators and question formats ($GCC\ binary\ SV = 0.010; LCC\ binary\ SV = 0.008; GCC\ percent\ SV = 0.009; LCC\ percent\ SV = 0.007$). Considering both bias and sampling variance simultaneously, we find that the LCC percent estimator performs the best and that the percent question form has slightly lower error ($GCC\ binary\ RMSE = 0.102; LCC\ binary\ RMSE = 0.092; GCC\ percent\ RMSE = 0.097; LCC\ percent\ RMSE = 0.083$).

We next examine the distribution of estimates in realistic RDS samples and what features of RDS lead to performance deterioration compared to the RWS baseline scenario. Figure 4 shows that in the Project 90 network the GCC estimated using the binary question format performs poorly in each of the RDS scenarios, underestimating the population parameter substantially (GCC binary bias by scenario is $RDS\ baseline = -0.132, +imperfect = -0.127, +preferences = -0.130$, and $+misreporting = -0.067$). Underestimation begins with the RDS baseline scenario and persists, which indicates that problems for this estimator arise from the use of multiple seeds, convenience seed selection, without replacement design, and/or branching. Because we do not see comparable biases in the percent format under these scenarios (GCC percent bias by scenario is $RDS\ baseline = -0.010, +imperfect = -0.007, +preferences = -0.008$, and $+misreporting = -0.006$), we attribute this bias to the binary question format's restrictions on effective sample size because the binary question format is only asked of non-seed respondents who recruit others, while the percent format can be asked of any non-seed sample participant.

**Figure 4. Performance of Hardiman Katzir estimators by estimator and question format in RWS and RDS scenarios on the Project 90 data set.**

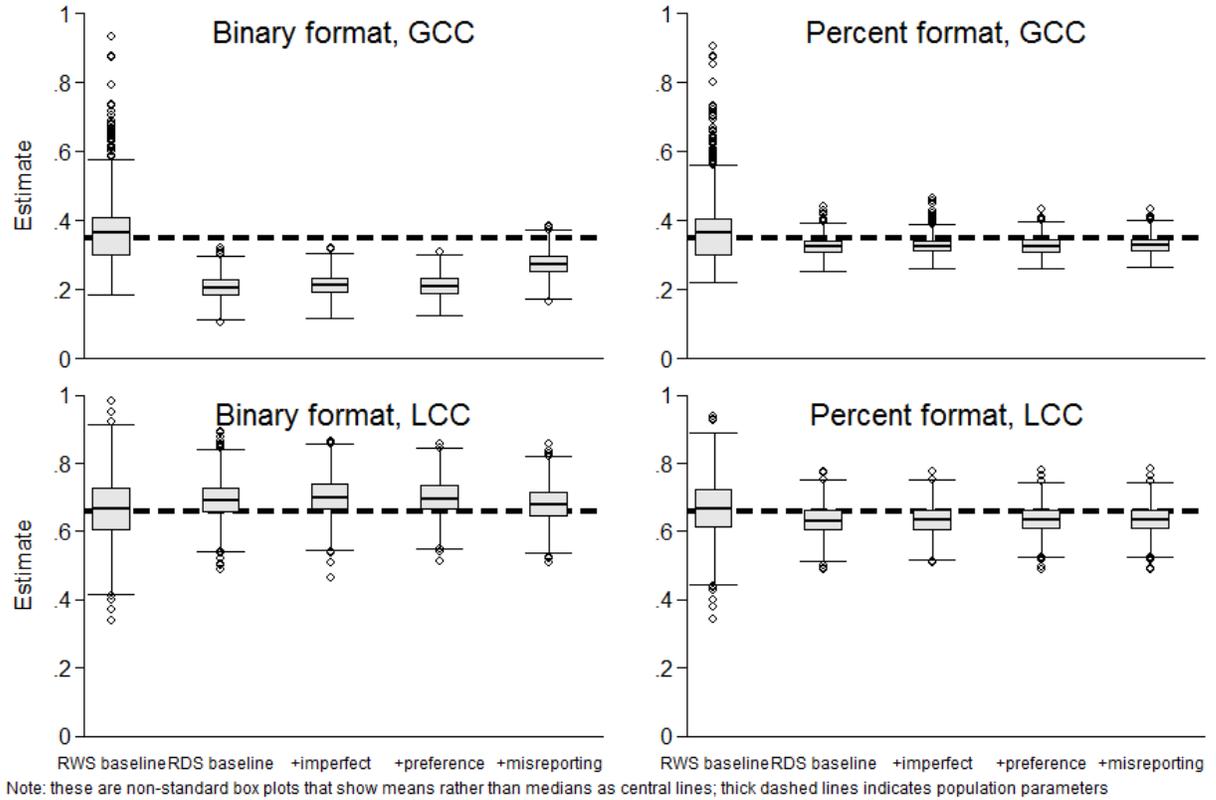

Note: these are non-standard box plots that show means rather than medians as central lines; thick dashed lines indicates population parameters

The LCC estimators perform well in Figure 4. The binary question format of the LCC slightly overestimates clustering (LCC binary bias by scenario is $RDS\ baseline = 0.039$, $+imperfect = 0.044$, $+preferences = 0.038$, and $+misreporting = 0.019$), while the percent form slightly underestimates it (LCC percent bias by scenario is $RDS\ baseline = -0.019$, $+imperfect = -0.016$, $+preferences = -0.016$, and $+misreporting = -0.015$).

Estimates obtained in all RDS scenarios in the Project 90 network exhibit low sampling variance (ranging from 0.001 to 0.003), substantially lower than was found for the RWS scenarios. This result follows from RDS's without replacement design, which tends to yield lower sampling variance than RWS's with replacement design. RMSEs in the worst case scenarios, which contain all RDS deviations from RWS that we examine, are lower than we found for the RWS baseline scenarios in all cases: in the +misreporting scenarios, RMSEs are $GCC\ binary\ RMSE = 0.076$; $LCC\ binary\ RMSE = 0.057$; $GCC\ percent\ RMSE = 0.034$; $LCC\ percent\ RMSE = 0.045$.

**Table 4. Distributions of absolute bias statistics and RMSEs in the 29 Facebook networks studied, by scenario, estimator, and question format.**

|  | Absolute Bias | | | | RMSE | | | |
|---|---|---|---|---|---|---|---|---|
|  | GCC | | LCC | | GCC | | LCC | |
|  | Binary | Percent | Binary | Percent | Binary | Percent | Binary | Percent |
| RWS baseline | | | | | | | | |
| Min | 0.000 | 0.000 | 0.000 | 0.000 | 0.019 | 0.006 | 0.040 | 0.023 |
| 25th percentile | 0.000 | 0.000 | 0.001 | 0.000 | 0.022 | 0.008 | 0.046 | 0.028 |
| Median | 0.001 | 0.000 | 0.001 | 0.001 | 0.023 | 0.008 | 0.048 | 0.030 |
| 75th percentile | 0.001 | 0.000 | 0.002 | 0.001 | 0.024 | 0.009 | 0.052 | 0.032 |
| Max | 0.002 | 0.000 | 0.005 | 0.003 | 0.027 | 0.012 | 0.058 | 0.040 |
| RDS baseline | | | | | | | | |
| Min | 0.012 | 0.006 | 0.002 | 0.000 | 0.025 | 0.010 | 0.051 | 0.025 |
| 25th percentile | 0.019 | 0.009 | 0.010 | 0.004 | 0.031 | 0.014 | 0.055 | 0.029 |
| Median | 0.020 | 0.011 | 0.012 | 0.007 | 0.033 | 0.016 | 0.057 | 0.033 |
| 75th percentile | 0.026 | 0.013 | 0.017 | 0.009 | 0.037 | 0.020 | 0.062 | 0.038 |
| Max | 0.041 | 0.021 | 0.025 | 0.015 | 0.051 | 0.028 | 0.074 | 0.052 |
| RDS misreporting | | | | | | | | |
| Min | 0.030 | 0.002 | 0.032 | 0.001 | 0.043 | 0.009 | 0.064 | 0.025 |
| 25th percentile | 0.046 | 0.006 | 0.044 | 0.003 | 0.055 | 0.012 | 0.068 | 0.028 |
| Median | 0.050 | 0.008 | 0.048 | 0.005 | 0.057 | 0.014 | 0.071 | 0.031 |
| 75th percentile | 0.054 | 0.010 | 0.051 | 0.007 | 0.061 | 0.017 | 0.074 | 0.037 |
| Max | 0.065 | 0.016 | 0.061 | 0.014 | 0.070 | 0.024 | 0.089 | 0.055 |

We next turn to results in the Facebook networks. Table 4 shows how absolute values of bias ("absolute bias") and RMSEs are distributed within these networks by estimator and question format in three focal scenarios (RWS baseline, RDS baseline, and RDS misreporting). We display these scenarios because the +imperfect and +preferences scenarios made little difference in the results. We do not show the low sampling variance we found in all scenarios for the Facebook networks (a maximum of 0.004 across networks in any scenario). The estimators exhibit almost no bias in the RWS baseline scenarios, with a maximum that is substantially lower than was seen in the Project 90 network. The RWS baseline scenario also tends to produce much lower RMSEs in these networks than it did in the Project 90 network.

The RDS scenarios also yield lower bias in the Facebook networks than they did in the Project 90 network, with maximum observed values all lower in these networks. In terms of bias, the Facebook networks indicate that the binary measures are the most biased, with the LCC being less biased than the GCC. The Facebook networks also have lower RMSEs than the Project 90 network. In terms of RMSEs in the realistic RDS scenarios, results from the Facebook networks suggest that the percent question format is preferable to the binary format and that the GCC is slightly preferred over the LCC after accounting for sampling variance (recall that the LCC had lower bias). In total, median RMSEs observed in the RDS scenarios in the Facebook networks are only slightly larger than the median RMSEs obtained in the RWS baseline scenarios, which indicates that the clustering coefficient estimators maintain reasonable properties for application to RDS samples.

**Application of Data Collection Instruments in Six Empirical Surveys**

**Table 5. Summary of Item Response Rates for Clustering Questions in Empirical Surveys.**

| Survey location | Population | Format | Reports[a] | Invalid % | Mean of valid |
|---|---|---|---|---|---|
| Shanghai, China | FSW[b] | Percent | 515 | 0.0% | 23.2% |
| Liuzhou, China | FSW[b] | Percent | 576 | 0.5% | 42.3% |
| Cebu, Philippines | PWID[c] | Binary | 380 | 14.2% | 78.7% |
| Mandaue, Philippines | PWID[c] | Binary | 291 | 8.3% | 91.7% |
| Ottawa, Canada | PWID[c] | Percent[e] | 364 | 11.5% | 67.0% |
| La Plata, Argentina | Veg[d] | Percent | 145 | 5.5% | 32.0% |
| La Plata, Argentina | Veg[d] | Binary | 131 | 36.6% | 30.1% |

**Notes: a-We refer to reports rather than sample size because for the binary questions some respondents report on multiple relationships; b-FSW stands for female sex workers; c-PWID stands for persons who inject drugs; d-Veg stands for self-identifying vegetarians and vegans; e-The format used in the Ottawa Study is an interaction grid in which respondents identify which peers know one another, see appendix.**

We now discuss six empirical RDS surveys collected in diverse hidden populations in multiple countries by different research teams that asked respondents the types of questions needed to estimate network clustering. Two studies examine female sex workers in China, two examine people who inject drugs in the Philippines, one looked at people who inject drugs in Canada, and the last survey, which contained both of our proposed question formats, looked at vegetarians and vegans in Argentina. For the sake of brevity, we omit full descriptions of these studies in the main text but provide complete details in Appendix B. We focus on the proportion of invalid item responses ("Invalid %") in each survey across question formats, where we define invalid responses as cases where respondents did not answer the question, gave responses of "don't know", or otherwise offered evidence that they did not understand or wish to answer the question. We also compare the mean values of valid responses ("Mean of valid") between relevant survey pairs (comparing the two surveys in China to each other, and the two surveys in The Philippines to each other), and within individuals who answered both types of questions in the survey in Argentina.

Table 5 summarizes the item response patterns in these empirical surveys. Respondents were much more likely to give invalid responses to the binary question format than to the percent question format. More speculatively, we can make some claims about conceptual validity by examining the cross-site concordance in the means of valid responses within the two sets of paired surveys. For instance, the means of valid responses in the female sex worker surveys collected by overlapping research teams in two cities in China are moderate (23.2%-42.3%), while means of valid responses for the two surveys of persons who inject drugs in Philippines cities are much higher (78.7%-91.7%). We take these findings to indicate that the survey questions are measuring consistent phenomena. In addition, we find nearly identical means of valid responses between the two question formats implemented in the Argentina survey. Here, both the percent and binary measures find raw clustering levels in the 30.1-32.0% range, and we found that the respondent-specific average of binary format vs. percent format reports had a spearman correlation of 0.445, while the item-specific reports with potentially multiple binary reports per respondent had a polyserial correlation of 0.376. These correlations suggest a reasonably high level of agreement between question formats, even in the face of large amounts of missing data. Taken together, these results indicate that the questions tap into valid concepts, but they add another reason that researchers should prioritize implementing the percent question format: respondents seem more willing or able to answer it.

**Discussion and Conclusion**

Sociological interest in marginalized populations means researchers often confront situations where traditional sampling methods cannot be used. In such situations, RDS's peer-driven recruitment procedures yield large and diverse samples quickly and cheaply while maintaining respondent anonymity, which is why researchers have used it to sample hundreds of stigmatized, sensitive, and hidden groups. Prior methodological research on RDS has focused on its estimators of the population mean and avoided examining other features of hidden populations that it can reveal (with a few notable exceptions: Crawford 2016; Wejnert 2010). This avoidance is strategic: practical considerations limit researchers' ability to uncover most features of the underlying population social network. In this paper, we developed recent work in computer science and proposed new data collection protocols and estimators that allow researchers to examine one network feature of broad interest, clustering. We began by considering estimators of network clustering proposed for random walk sampling (RWS) and expanded their application to the case of RDS, with careful attention to practical differences between RDS and RWS. We offered data collection protocols in the form of two different question formats that RDS surveys could adopt in the field to estimate network clustering and studied their performance in simulations and implementation challenges in six empirical surveys.

Overall, we recommend that researchers using RDS surveys begin asking respondents the types of questions that would allow for clustering coefficient estimation. While RDS estimators of the population mean often fail in the face of unmet assumptions about sample recruitment (Gile and Handcock 2010; Verdery et al. 2016; Merli, Moody, Smith, et al. 2015; Lu et al. 2013; Lu et al. 2012; Goel and Salganik 2010; Tomas and Gile 2011; McCreesh et al. 2012), we find that clustering coefficient estimators perform well even when core RDS assumptions are violated. We found that the percent question format can be asked of more respondents, yielded better results in a simulation study, and appeared to be better understood by respondents in empirical studies. The two clustering estimators perform similarly, but the GCC estimator had lower total errors than the LCC estimator in most networks we studied. However, sampling variance's contribution to RMSE drives this result, so researchers concerned about bias may prefer to stick to the LCC estimator, which we found tends to exhibit lower bias.

We hope that methods for estimating clustering coefficients from RDS data will spur additional substantive and methodological contributions. Substantively, clustering is a core property that distinguishes human social networks from random graphs (Watts and Strogatz 1998). Structural hypotheses about network diffusion derived from mathematical models hold that levels of clustering influence diffusion dynamics at the network level. For example, such models suggest that *ceteris paribus* moving from low to moderate clustering of the risk network increases transmission (Keeling and Eames 2005), but moving from moderate to high clustering does not change transmission substantially until very high levels when the network becomes disconnected (Newman 2003). Using clustering coefficients from RDS data could allow researchers to confirm the insights of these mathematical models of network structure and disease diffusion with macro-comparative methods.[2] Second, clustering in the social network may be associated with differences in risk behaviors like unprotected sex at the individual level. Prior research finds that an individual's network clustering interacted with the density of contraceptive users strongly affects fertility control (Kohler, Behrman, and Watkins 2001), but that such normative reinforcement can also facilitate the spread of unhealthy behaviors (Yamanis et al. 2015). Previous studies of this topic have been limited to traditional survey populations, however, and the approaches developed in this paper will enable researchers to test these hypotheses in a more diverse series of hidden populations.

In addition, estimators of network clustering can offer methodological improvements to RDS. An extension could yield additional validation of promising variants of RDS mean estimators that use exponential random graph modeling and algorithmic simulation to obtain less biased, lower variance results (Gile and Handcock 2011). Currently, these approaches model clustering as a byproduct of dyadic homophily. With empirical estimates of clustering, researchers using such algorithms could confirm the clustering coefficients produced in their algorithms. A second contribution could allow researchers to test one of the most central but least often evaluated assumptions of RDS, that the network contains a "giant component" where the vast majority of people are reachable through chains of arbitrary length through the network ties (Volz and

---

[2] For clarity in this example, we assume that the social network that the RDS chain traverses is a close proxy for the risk network for the disease, a connection that future research should examine more closely.

Heckathorn 2008). Using random graph methods from the physics and computer science traditions that generate network structures from degree distributions and clustering coefficients (Newman, Strogatz, and Watts 2001; Heath and Parikh 2011), researchers may also be able to determine if they are sampling a network with "bottlenecks," i.e., where there are few links between cohesive groups in the network, a feature which many in the RDS community link to poor estimate quality (Toledo et al. 2011). This would add to the emerging diagnostic toolkit being developed for RDS (Gile, Johnston, and Salganik 2015). A related extension of this approach could calculate the "structural risk" of a network sampled with RDS by applying percolation or other diffusion models to examine the size and speed of hypothetical epidemics spreading on the modeled network (Britton et al. 2008; Merli, Moody, Mendelsohn, et al. 2015) – a potential early warning system of a given hidden population's epidemic potential gathered directly from RDS.

Such extensions and future directions lie outside of the scope of the present article. However, we emphasize that rather than an end point, we view this as a beginning. The benefits from estimating clustering in RDS samples are large, and we encourage researchers to begin deploying survey questions needed for their calculation. In either case, further attention to RDS's ability to tell us more about hidden populations than disease prevalence is an important next step for the literature to take.

**Appendix A: Other Seed Selection Procedures**

In the main text of the article we defined all of the RDS scenarios as starting from a uniform random sample of seeds. In this appendix, we consider 4 alternative scenarios in the Project 90 network that vary seed selection procedures but otherwise retain all features of the "+misreporting" scenarios (we found no difference for the other RDS scenarios but do not report on them here). In these scenarios we select seeds: 1) uniformly at random from white nodes only ("+white"); 2) uniformly at random from non-white nodes only ("+non-white"); 3) with probability proportional to their level of local clustering ("+high cluster); 4) with probability inverse proportional to their level of local clustering ("+low cluster).

Table A1 shows the results under these alternative seed selection scenarios. We found few meaningful differences between the results provided in the main text of the manuscript and those obtained with alternative seed selection procedures. None of the biases changed directions, the largest change in the RMSEs was a level of 0.03 (for the GCC binary estimates), and, in general, the rank ordering of estimator performance was maintained with the percent question formats having lower RMSEs than the binary formats.

**Table A1. Bias and RMSEs in the Project 90 network, by alternative seed selection scenario, estimator, and question format.**

|  | Bias Measures | | | | RMSE | | | |
|---|---|---|---|---|---|---|---|---|
|  | GCC | | LCC | | GCC | | LCC | |
| Scenario | Binary | Percent | Binary | Percent | Binary | Percent | Binary | Percent |
| +misreporting | -0.067 | -0.006 | 0.019 | -0.015 | 0.076 | 0.034 | 0.057 | 0.045 |
|  |  |  |  |  |  |  |  |  |
| +non-white seeds | -0.097 | -0.038 | 0.006 | -0.040 | 0.102 | 0.042 | 0.053 | 0.057 |
| +white seeds | -0.067 | -0.006 | 0.019 | -0.016 | 0.076 | 0.033 | 0.057 | 0.045 |
| +high cluster seeds | -0.076 | -0.014 | 0.010 | -0.016 | 0.085 | 0.033 | 0.056 | 0.045 |
| +low cluster seeds | -0.077 | -0.030 | 0.043 | -0.037 | 0.085 | 0.038 | 0.067 | 0.057 |

**Appendix B: Survey Questions Used in Empirical Surveys**

This appendix provides the specific survey questions used in the six empirical studies reviewed in the "Applications in Empirical Surveys" section.

The Shanghai Women's Health Study was collected in 2007 using RDS of female sex workers living in Shanghai, China (Merli et al. 2010; Yamanis et al. 2013). This study's protocol was approved by the Research Ethics Committee of the University of Wisconsin, Madison and the Shanghai Institute of Planned Parenthood Research. This survey used a percent question format, where non-seed respondents were asked the following two questions:

> Q.901. In Shanghai, how many of this kind of sex workers do you know? You know how to address them, they know how to address you, and you have met or contacted them in the past month.
>
> Q.904. Among those people (the people in 901), how many do both you and your contact (the person who introduced you to the project) know?

We obtain the percent by dividing the answer to Q.904 by the answer to Q.901.

The RDS component of the PLACE-RDS Comparison Study sampled female sex workers in Liuzhou, China in 2010 (Weir et al. 2012). This study was approved by the Research Ethics Committee of the National Center for STD Control, China and the Institutional Review Boards at the University of North Carolina and Duke University. This survey was conducted by members of the same team as the Shanghai study, and it also used the percent format by asking two iterative questions. Non-seed respondents in this survey were asked:

> Q.901. In Liuzhou city (including Liuzhou counties), how many women do you know personally who are sex workers? By sex worker, I mean that they are paid money in exchange for sex. By know personally, I mean:
>
> > -you know their name and they know yours
> >
> > -you know who they are and they know you
> >
> > -you have seen or contacted them in the past four weeks
>
> Q.904. Of the (repeat response number from 901) sex workers you know, how many are also known by the person who gave you this coupon?

As above, we obtain the percentage by dividing the answers to these questions.

The Characterizing the Social Networks of Women and Men in Ottawa who Inject Drugs to Drive Prevention Programming Study sampled people who inject drugs in Ottawa, Canada in 2007 (Pilon et al. 2011). The Ottawa Hospital Research Ethics Board approved the study. This study asked respondents a percentage format of the question, but the approach used to collect these data differed from the format asked in the two studies of female sex workers in China that we reviewed above. Rather than asking respondents counts of potential recruitees that know the respondents' recruiter, trained interviewers directly asked respondents questions to elicit ego networks, and then asked them to complete an interaction grid recording contact between ego network peers. First respondents were asked to list members they know:

> Q.1.) First, please think back over the last 30 days about the people with whom you have had more than casual contact. These would be people that you have seen or have spoken to on a regular basis. Most of these close contacts would be people such as friends, family, sex partners, people you inject drugs with, or people you live with. Let's make a list of these people starting with those who inject drugs. Please use only initials, or some other identifier that will make sense to you such as a made up name. Please do not use their last names. We will use this list to make sure we know which individuals we are talking about. Remember that we are interested in people that you've had contact with in the last 30 days.

Then interviewers worked with respondents to fill out an interaction grid on the basis of the following instructions:

> Q3. (Following step 2, transfer the names of all the network members from the previous question onto the interaction grid – list the contacts in the ID column going down from 1-20. For each person listed, ask the subject to indicate which of the other individuals on the list that particular person knows or has contact with. Indicate whether they know one another by placing an X in the appropriate box. You are working down through the columns, not across. E.g., if Sam is ID#1, you will go down column 1 and ask if Sam knows Tom, Mary, Mac, OT, etc. In Column 1, you will end up with an X beside each of Sam's contacts. Next, move to column 2 and do the same for Tom, then move to column 3, column 4, etc.)

[Interaction grid: 20×20 matrix with ID column on left labeled 1-20 and column headers 1-20 across the top, lower triangular blank cells for entries]

The interaction grid gives us the egocentric networks (each individual participant's social network) and the card system gives us the socio-metric network).

We obtained percentages by calculating the ego-network density of this matrix. We leave it for future investigation to determine whether this approach provides meaningfully different results than the percent format question recommended in text, because implementing this interaction grid adds substantial time to the data collection process.

The third and fourth studies we examine come from two surveys that were part of the Integrated HIV Biological and Serological Surveillance Study fielded by researchers at the Philippines Department of Health in 2013 (National Epidemiology Center, Department of Health, Philippines 2014). Data collection was a surveillance activity and was not subject to institutional review board approval, but secondary data analysis received approval from the Institutional Review Board of the University of North Carolina at Chapel Hill. These studies surveyed people who inject drugs in Cebu City and Mandaue City, The Philippines, a binary format of the question. Specifically, they asked respondents the following question:

> 1. Does the person you gave a coupon to, and your recruiter (that is, the person who gave you your coupon) know each other?

Finally, we examine early results from a sixth RDS study. The pilot survey *EncuestaVeg* sampled vegetarians and vegans living in La Plata, Argentina, where avoiding meat is such a rare activity as to make those who identify with the practice a hidden population. This ongoing pilot survey was begun in June, 2016; we report on results obtained as of September, 2016. The protocol for this survey was approved by the Institutional Review Board of the Pennsylvania State University. In it, respondents were asked both the percent and the binary question. First, during the primary interview, non-seed respondents were asked a percent format question:

> 13.1. Think about all the people you know who live in the city of La Plata ages 18 and up. How many vegans and vegetarians total do you know (you know their name and they know yours)?

> 13.9. Think of the person who gave you the code. Of the rest of the vegans and vegetarians who you know in La Plata, how many also know the person who gave you the code?

Percentages were obtained by dividing these questions. Note that Q13.9 did not specifically reference the answer given for Q13.1, and also that the response entry was open ended. Some respondents said larger numbers in 13.9 than they did for 13.1, while others gave string responses such as *"todos* [all]", or "*Casi todos* [nearly all]". In the main text, we report these cases as invalid responses (except *todos*, which we code as 100%). In addition to the percent question format, recruiting participants in *EncuestaVeg* who returned to complete the follow-up survey were asked a series of questions about who they invited to participate and a question that allows us to calculate the binary question format. Specifically, for each person they invited, they were asked:

> Q.F.18. Does this person know the person who gave you the code to answer the survey?

We use answers to this question as the binary question format.